\documentclass[structabstract]{aa}

\usepackage{url}
\usepackage[breaklinks=true]{hyperref}
\usepackage{twoopt}
\usepackage[english]{babel}          

\usepackage{natbib}
\bibpunct{(}{)}{;}{a}{}{,} 

\usepackage{siunitx}

\usepackage[farskip=0pt]{subfig}

\usepackage{multirow,bigstrut,ctable}
\usepackage[varg]{txfonts} 
\usepackage{threeparttable}

\def \deg         {\text{$^{\circ}$}}
\def \arcmin      {\text{$^\prime$}}
\def \arcsec      {\text{$^{\prime\prime}$}}
\def \hour        {$^{\mathrm{h}}$}
\def \min         {$^{\mathrm{m}}$}
\def \sec         {$^{\mathrm{s}}$}

\def \mujybeam    {$\mathrm{\mu}$Jy\,beam$^{-1}$}

\def \mach	  {\mathcal{M}}

\newcommand{\Hi}{\text{H\textsc{i}}}
\newcommand{\hms}[3]{{#1}\hour{#2}\min{#3}\sec}
\newcommand{\dms}[3]{{#1}\deg{#2}\arcmin{#3}\arcsec}
\newcommand{\beam}[2]{{#1}\arcsec$\times${#2}\arcsec}

\newcommand{\Msun}{\text{$\rm M_\odot$}}

\def \target {Abell~3527-bis}

\begin{document}

\title{Tracing low-mass galaxy clusters using radio relics:\\ the discovery of Abell 3527-bis}
\titlerunning{Tracing galaxy clusters using radio relics}

\author{F.~de~Gasperin\inst{1}
\and H.~T.~Intema\inst{1}
\and J.~Ridl\inst{2}
\and M.~Salvato\inst{2}
\and R.~van~Weeren\inst{4}
\and \\A.~Bonafede\inst{3}
\and J.~Greiner\inst{2}
\and R.~Cassano\inst{5}
\and M.~Br\"uggen\inst{3}}

\authorrunning{F.~de~Gasperin et al.}

\institute{Leiden Observatory, Leiden University, P.O.Box 9513, NL-2300 RA, Leiden, The Netherlands, \\ \email{fdg@strw.leidenuniv.nl}
\and Max-Planck-Institut f\"ur extraterrestrische Physik, Giessenbachstrasse Postfach 1603, Garching D-85740, Germany
\and Universit\"at Hamburg, Hamburger Sternwarte, Gojenbergsweg 112, D-21029, Hamburg, Germany
\and Harvard-Smithsonian Center for Astrophysics, 60 Garden Street, Cambridge, MA 02138, USA
\and IRA-INAF, via P. Gobetti 101, 40129 Bologna, Italy}

\date{Received ... / Accepted ...}

\abstract
{Galaxy clusters undergo mergers that can generate extended radio sources called radio relics. Radio relics are the consequence of merger-induced shocks that propagate in the intra cluster medium (ICM).}
{In this paper we analyse the radio, optical and X-ray data from a candidate galaxy cluster that has been selected from the radio emission coming from a candidate radio relic detected in NRAO VLA Sky Survey (NVSS). Our aim is to clarify the nature of this source and prove that under certain conditions radio emission from radio relics can be used to trace relatively low-mass galaxy clusters.}
{We observed the candidate galaxy cluster with the Giant Meterwave Radio Telescope (GMRT) at three different frequencies. These datasets have been analysed together with archival data from ROSAT in the X-ray and with archival data from the Gamma-Ray Burst Optical/Near-Infrared Detector (GROND) telescope in four different optical bands.}
{We confirm the presence of a 1~Mpc long radio relic located in the outskirts of a previously unknown galaxy cluster. We confirm the presence of the galaxy cluster through dedicated optical observations and using archival X-ray data. Due to its proximity and similar redshift to a known Abell cluster, we named it \target{}. The galaxy cluster is amongst the least massive clusters known to host a radio relic.}
{We showed that radio relics can be effectively used to trace a subset of relatively low-mass galaxy clusters that might have gone undetected in X-ray or Sunyaev-Zel'dovich (SZ) surveys. This technique might be used in future deep, low-frequency surveys such as those carried on by the Low Frequency Array (LOFAR), the Upgraded GMRT (uGMRT) and, ultimately, the Square Kilometre Array (SKA).}

\keywords{galaxies: clusters: individual: \target{} -- large-scale structure of Universe -- radio continuum: general}

\maketitle

\section{Introduction}
\label{sec:introduction}
Cosmic structure forms hierarchically, in a bottom-up scenario, with smaller structures merging to form bigger ones. On the largest scales, clusters of galaxies merge releasing energies on the order of $10^{64}$ erg on time scales of 1--2 Gyr \citep[e.g.][]{Hoeft2007}. During these events, large-scale shock waves with moderate Mach numbers ($\lesssim 4$) are created in the intra cluster medium (ICM). Shocks are collision-less features whose interactions in the hot plasma are mediated by electromagnetic fields. They act on the ICM accelerating a fraction of the thermal distribution of particles transforming them into non-thermal populations of cosmic rays (CRs) as initially proposed by \cite{Ensslin1998}.

The mechanism initially thought to be able to accelerate CRs was diffusive shock acceleration \citep[DSA, e.g.][]{Drury1983}, however the efficiency of acceleration via the DSA mechanism in weak shocks is thought to be very low \citep{Malkov1995}. A possibility to overcome this problem is that shocks re-accelerate populations of already mildly relativistic electrons, instead of directly accelerating them from the thermal pool \citep[e.g.][]{Markevitch2005,Kang2011,Kang2012,Kang2016}. Some observational pieces of evidence for this scenario are available \citep{Bonafede2014a,Shimwell2015} but are not conclusive. Recently, particle-in-cell (PIC) simulations provided new insight into proton and electron acceleration in weak shocks in galaxy clusters. They indicate that shock drift acceleration (SDA) might be efficient in injecting particles into the DSA process \citep{Caprioli2014,Guo2014}.

Regardless of the acceleration method, the empirical evidence of the presence of large-scale shocks in the ICM is given by radio relics\footnote{We use the commonly accepted name ``radio relic'' while other authors prefer the name radio gischt.} \citep[for a review see e.g.][]{Feretti2012,Brunetti2014}. These are extended radio sources in the periphery of galaxy clusters. Their spectrum is rather steep ($\alpha \lesssim -1$)\footnote{$S_\nu \propto \nu^\alpha$; where $S_\nu$ is the flux density at frequency $\nu$.} and their emission is strongly polarised, with an ordered magnetic field usually aligned with the relic extension \citep[e.g.][]{vanWeeren2010a,deGasperin2015a}.

\begin{figure*}[!ht]
\centering
\subfloat[GMRT 148 MHz]{\includegraphics[width=.49\textwidth]{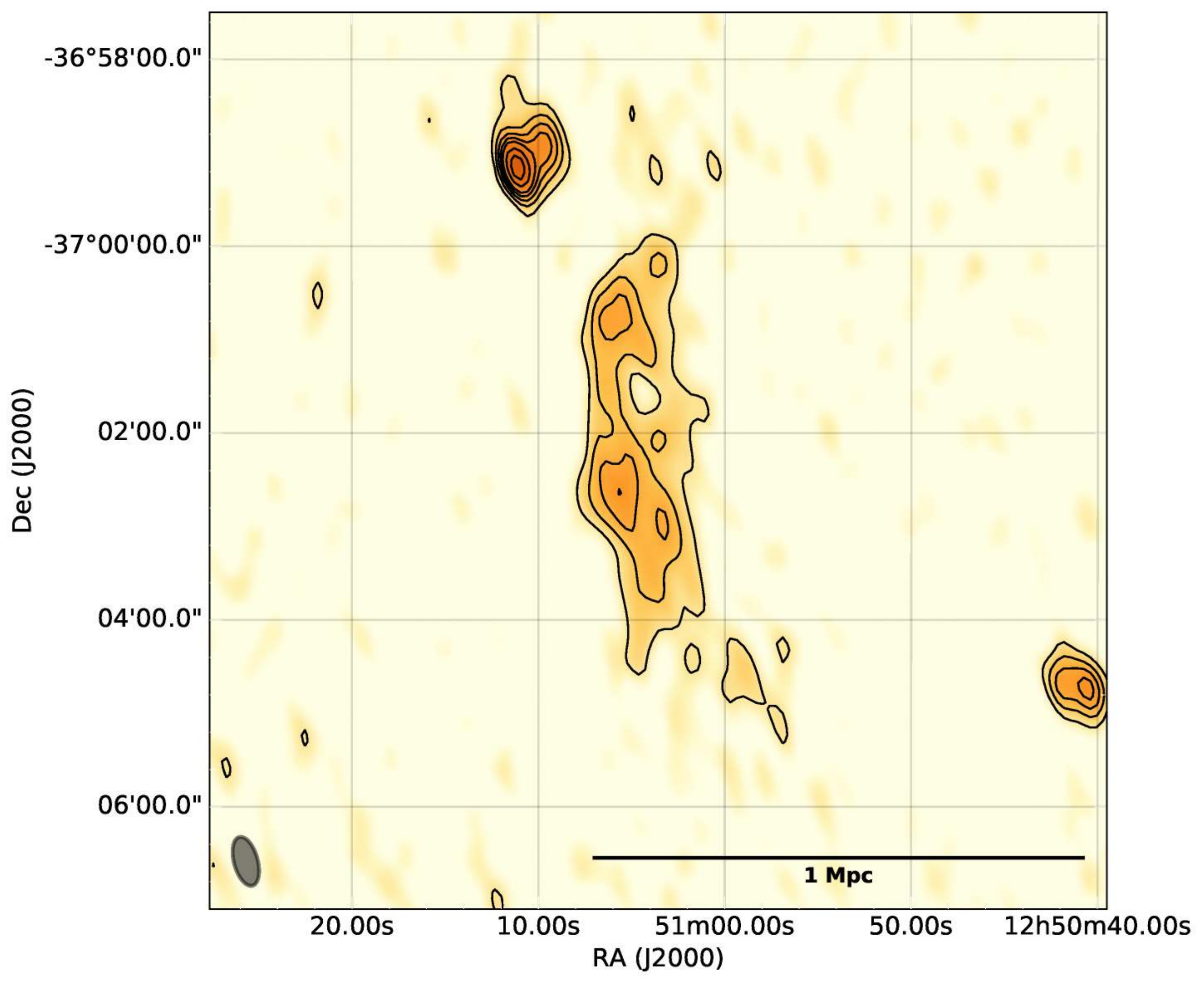}\label{fig:150}}
\subfloat[GMRT 323 MHz]{\includegraphics[width=.49\textwidth]{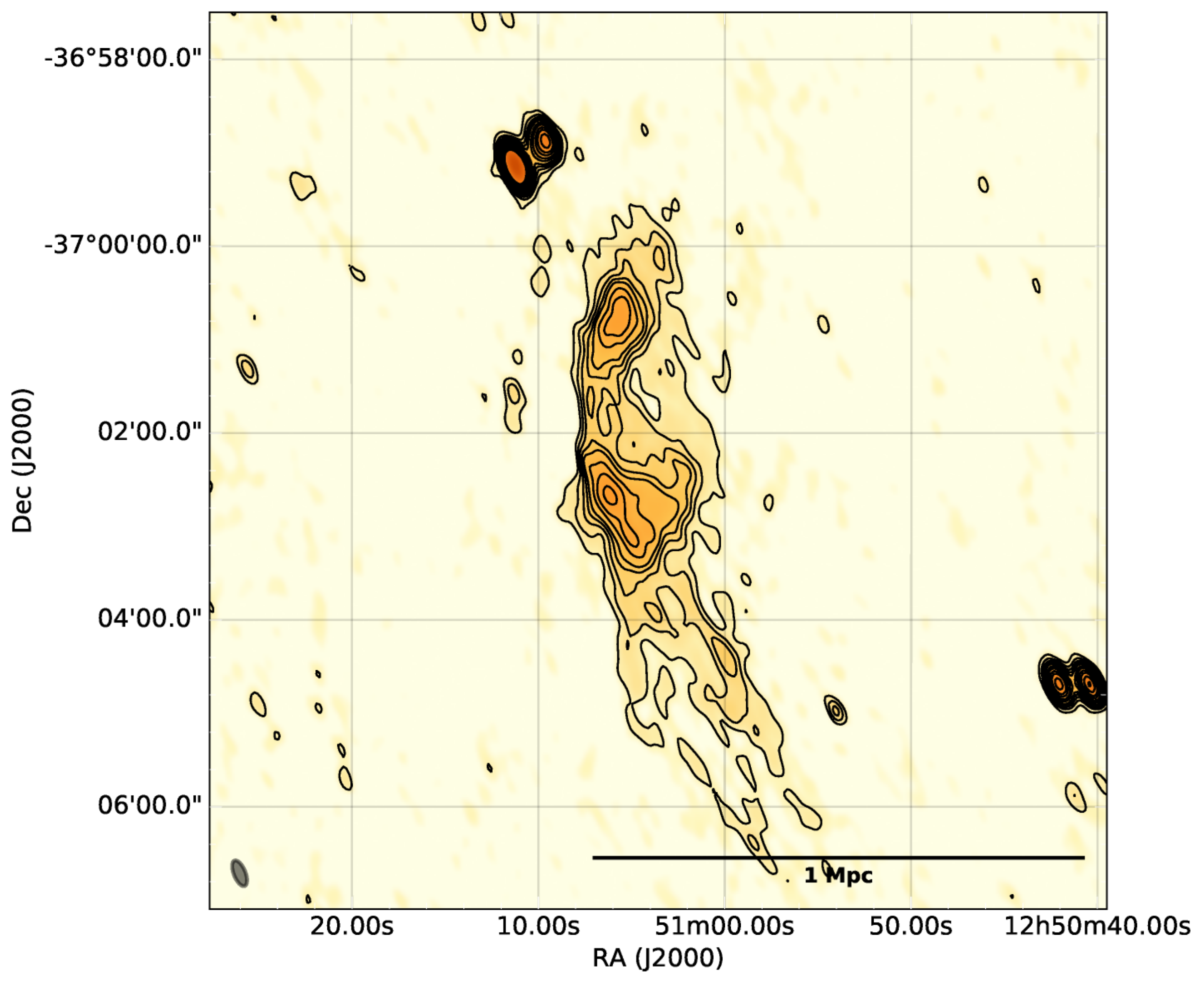}\label{fig:325}}\\
\subfloat[GMRT 608 MHz]{\includegraphics[width=.49\textwidth]{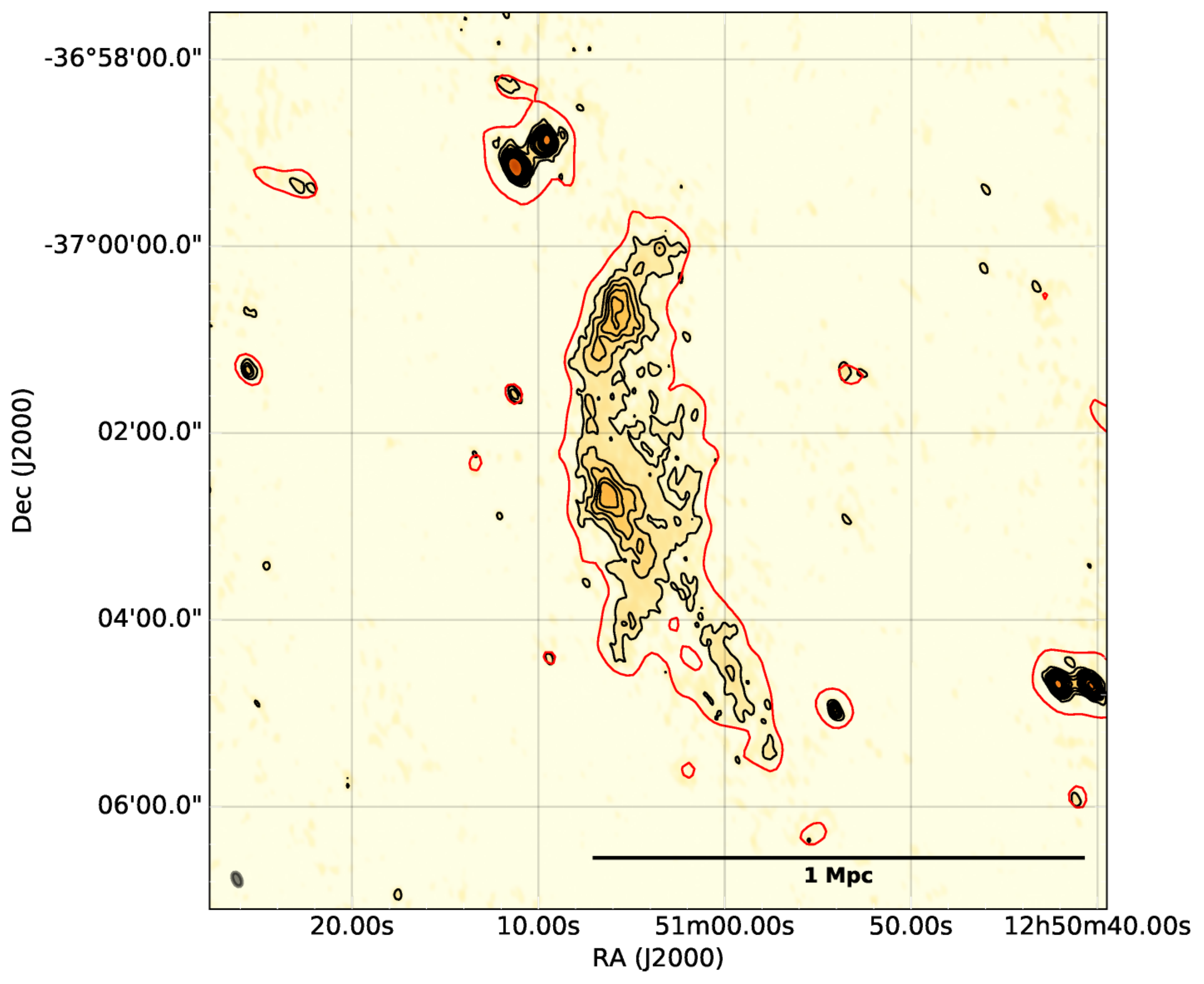}\label{fig:610}}
\subfloat[Spectral index (323-608 MHz)]{\includegraphics[width=.49\textwidth]{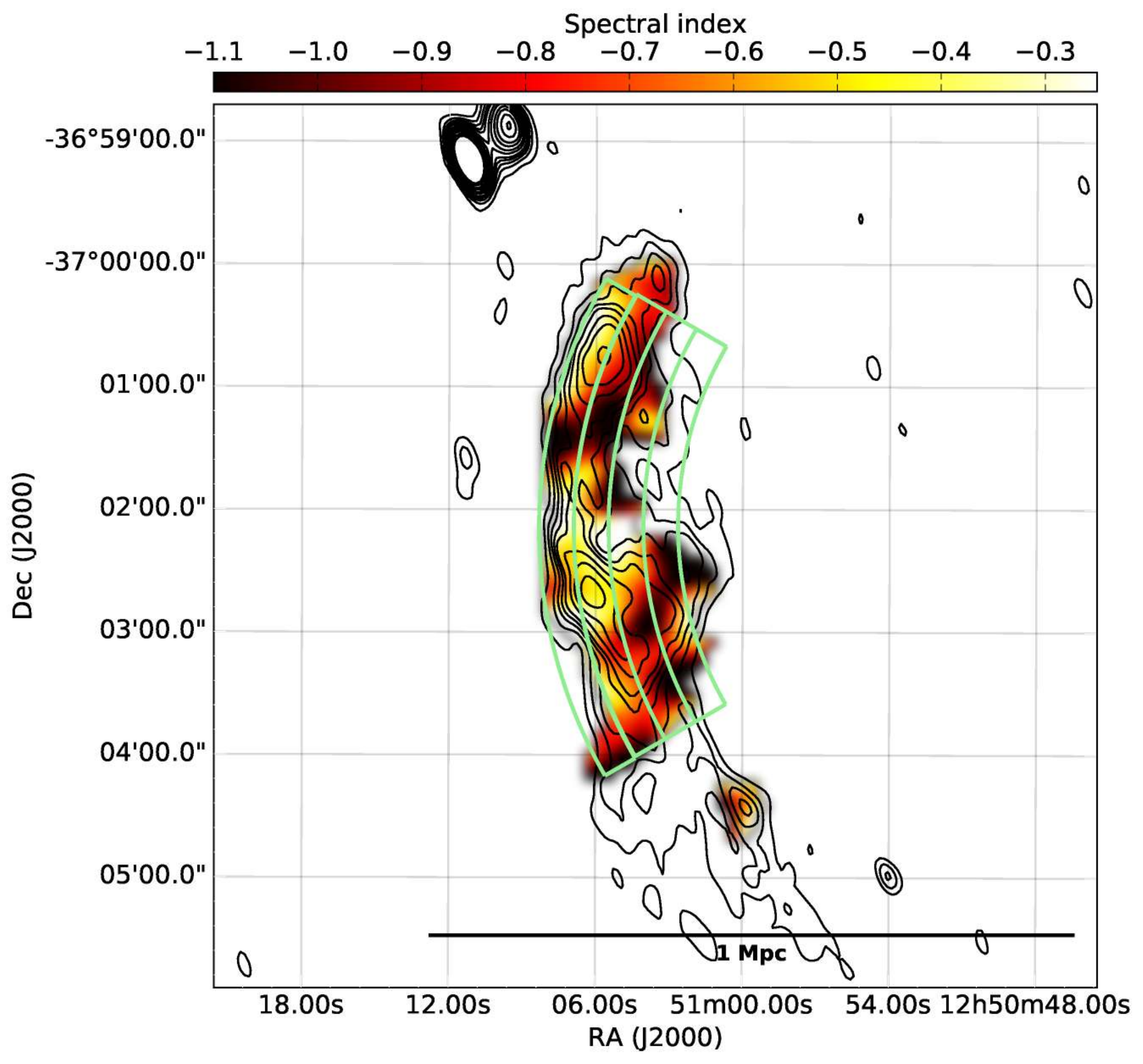}\label{fig:spidx}}
\caption{Radio image of the radio relic found in \target{} at 148, 323 and 608 MHz obtained with the GMRT. Contours are at $(1,2,3,4,5,7,9,12,15,20,25) \times 3\sigma$ with $\sigma=1400$, 67 and 44~\mujybeam{}, respectively. The resolutions of the three images are: \beam{32}{15} (148 MHz), \beam{18}{8} (323 MHz), and \beam{9}{5} (608 MHz). In the 608 MHz image the low-resolution (\beam{19}{15}) contours at $3\sigma$ ($\sigma=81$~\mujybeam) are over-plotted in red. The last panel shows the spectral index map obtained as described in the text. Over-plotted in green are the regions used to calculate the spectral index in Fig.~\ref{fig:spidx-plot}.}\label{fig:radio}
\end{figure*}

Thanks to the new SZ measurements obtained with the Planck satellite and the South Pole Telescope, many new galaxy clusters were recently identified \citep{PlanckCollaboration2015,Bleem2015}. However, SZ measurements are biased towards detecting massive clusters. The other traditional method to locate galaxy clusters is through their Bremsstrahlung radiation from the hot ICM \citep[e.g.][]{Boehringer1995,Ebeling1998a}, which is also biased towards massive and cool-core systems \citep{Eckert2011}. Given the correlation between radio-relic power and hosting-cluster mass \citep{deGasperin2014c}, most of the radio relics discovered so far are located in quite massive galaxy clusters ($M_{500} > {\rm few} \times 10^{14}$~\Msun).

Radio surveys can be used to detect previously unknown massive galaxy clusters by tracing powerful radio relics \citep{VanWeeren2012e}. \cite{Macario2014} also discovered a galaxy cluster thanks to the presence of extended radio emission, however the radio source could not be firmly classified. Less massive clusters has been also discovered by tracing diffuse radio emission in surveys \citep{Brown2011}.  The use of radio relics to trace low-mass galaxy clusters is a promising technique in the light of new high-sensitivity, large field-of-view telescopes such as LOFAR, uGMRT and, ultimately, SKA.

In this paper we report the discovery of a $\gtrsim 1$~Mpc long radio relic in the periphery of a previously unknown galaxy cluster. The presence of the galaxy cluster has been confirmed by optical and X-ray observations. Throughout the paper we adopt a fiducial $\Lambda$CDM cosmology with $H_0 = 70\rm\ km\ s^{-1}\ Mpc^{-1}$, $\Omega_m = 0.3$ and $\Omega_\Lambda = 0.7$. At the redshift of the target ($z\simeq0.2$) 1\arcsec = 3.3 kpc. Unless otherwise specified errors are at $1\sigma$.

\section{Observations}
\label{sec:observations}

We initially noticed the presence of this galaxy cluster due to the presence of extended radio emission in the NVSS \citep{Condon1998} and SUMSS \citep{Mauch2003} surveys. We found the radio source to be closely located to an X-ray excess listed in the ROSAT faint sources catalogue \citep[RXSJ125029.7-370220][]{Voges2000}. Despite being classified as point-like, the X-ray emission appears elongated in the SE-NW direction, while the radio emission, on the Eastern side of the cluster, has a N-S elongation. We interpreted this association as a previously unknown galaxy cluster associated to a radio relic. To confirm this claim we performed follow-ups in the radio. Furthermore, to confirm the presence of a cluster, we used optical data taken in 2014 as part of the programme for the follow up of ROSAT point-like sources with Gamma-Ray Burst Optical/Near-Infrared Detector (GROND)\footnote{PI Salvato, 093.A-9026(A).}.

\subsection{Radio}
\label{sec:observations-radio}

We performed GMRT observations of the candidate radio relic at three separate frequencies of 148, 323, and 608 MHz. The observations were taken on 8 s integration time, and the bandwidth was divided into 512 channels. Specific information on the observations is given in Table~\ref{tab:obs}. At the beginning and at the end of each observation 3C\,286 was observed and used as flux calibrator. The data reduction strategy included a direction-dependent calibration of ionospheric delay at all the three frequencies. The processing pipeline is described in detail in \cite{Intema2016}. The final image obtained from the lowest frequency observation (148 MHz) has been used as a model for the 323 MHz datasets, and the 323 MHz final image has been used as a model for the 608 MHz dataset. The calibrated dataset has been imaged using the CASA task CLEAN with robust=0 weighting and at a variety of resolutions applying different tapering to the $uv$-data in order to enhance the emission at specific scales. Images with uniform weighting were also produced to investigate the presence of compact sources within the radio-relic structure and none were found. Finally, the datasets at 323 and 608 MHz were used to produce a spectral index map (Fig.~\ref{fig:spidx}). In this case the datasets were imaged with uniform weighting, cutting the $uv$-coverage to have a common minimum value in wavelengths and tapering the data to obtain a common resolution. All images were primary-beam corrected.

\begin{table}
\centering
\begin{threeparttable}
\begin{tabular}{lccc}
\hline
Frequency (MHz)       & 148  & 323 & 608 \\
Obs date              & 29 Jan 2016 & 30 Jan 2016 & 15 Feb 2016 \\
Integration time (h)  & 5 & 6 & 6 \\
Rms noise (\mujybeam) & 1600 & 67 & 44 \\
Resolution            & \beam{32}{15} & \beam{18}{8} & \beam{9}{5} \\
Total bandwidth (MHz) & 16.66 & 33.33 & 33.33 \\
\hline
\end{tabular}
\end{threeparttable}
\caption{GMRT radio observations}\label{tab:obs}
\end{table}

\subsection{X-ray}
\label{sec:observations-x}

The only X-ray data available of \target{} is from the ROSAT all sky survey. We used the number count published in \cite{boller2016} to extract the source flux density. We note that the detection is based on only 28 counts and on an exposure time of 313.3 s. The source flux density has been calculated using webPIMMS\footnote{\url{https://heasarc.gsfc.nasa.gov/cgi-bin/Tools/w3pimms/w3pimms.pl}} and assuming: (i) an average galactic \Hi{} column density of ${nH} = 6.1\times 10^{20}$~cm$^{-2}$ \citep{Kalberla2005a}, (ii) a temperature of the ICM of $T = 4.4$~kev, (iii) a metallicity of 0.2 solar abundance. We obtained a flux density of $S_{[0.5-2\ {\rm kev}]} = (1.1 \pm 0.2) \times 10^{-12}$~ergs cm$^{-2}$ s$^{-1}$.

\begin{figure}
\centering
\includegraphics[width=.49\textwidth]{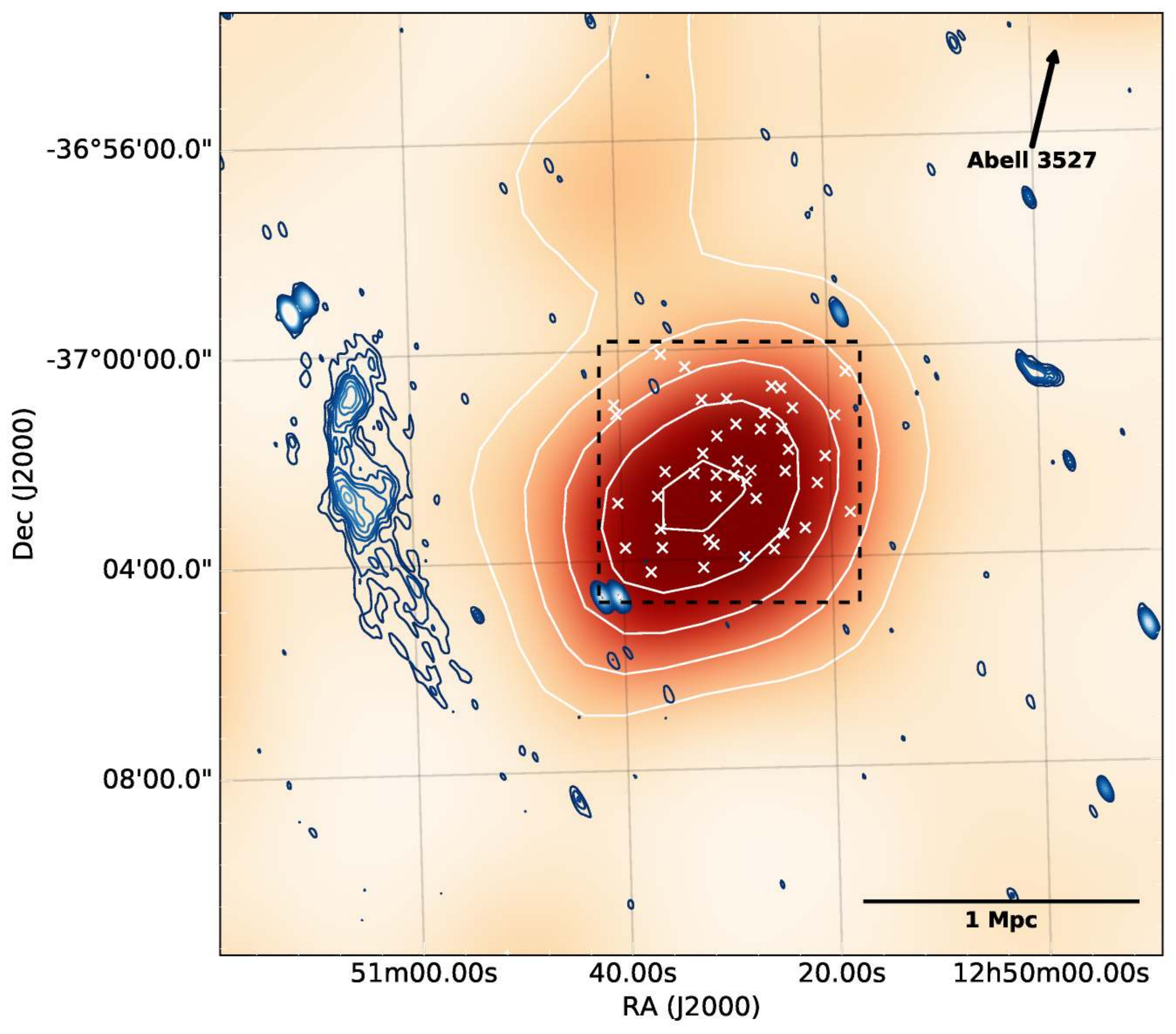}
\caption{Smoothed ROSAT image of \target{}. Over-plotted is the contour of the radio emission at 327 MHz, same contour level as in Fig.~\ref{fig:325}. No counter-relic is detected. The direction of the position of the nearby cluster Abell~3527 is labelled. Abell~3527 is 20\arcmin{} (4 Mpc at $z=0.2$) away from \target{}. White X show the position of optically-identified cluster members. The dashed line shows the field of view of the optical observation.}\label{fig:rosat}
\end{figure}

\subsection{Optical imaging}
\label{sec:observations-optical}

We retrieved GROND archival data, centred at the position of the ROSAT coordinated, observed on 8 Mar 2014. GROND was primarily designed to provide rapid multi-wavelength observations of gamma-ray bursts. It is a seven channel imager on the MPG 2.2 m telescope at La Silla/ESO, allowing for simultaneous imaging in the Sloan $g'r'i'z'$ and near-infrared $JHK$ bands with a field-of-view of $5.4'\times 5.4'$ ($10'\times10'$) and a pixel scale of $0.158''$ ($0.6''$) respectively. For our analysis we only used the optical bands since they bracket the 4000~\AA{} break, which is the most important spectral feature for determining the redshift. A comprehensive description of the instrument is given in \cite{Greiner2008}.

\begin{figure*}[!ht]
\centering
\includegraphics[width=.48\textwidth]{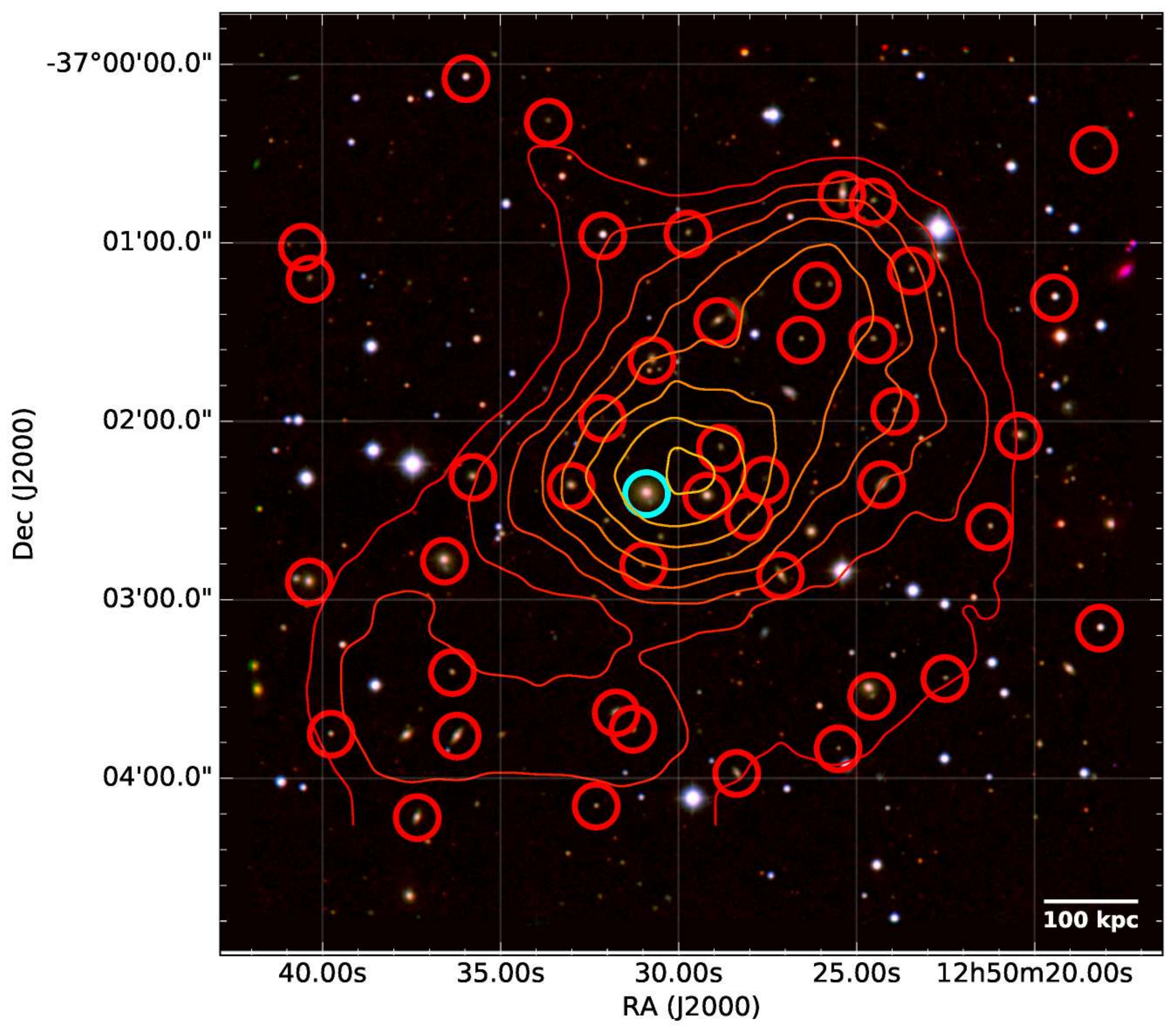}
\includegraphics[width=.48\textwidth]{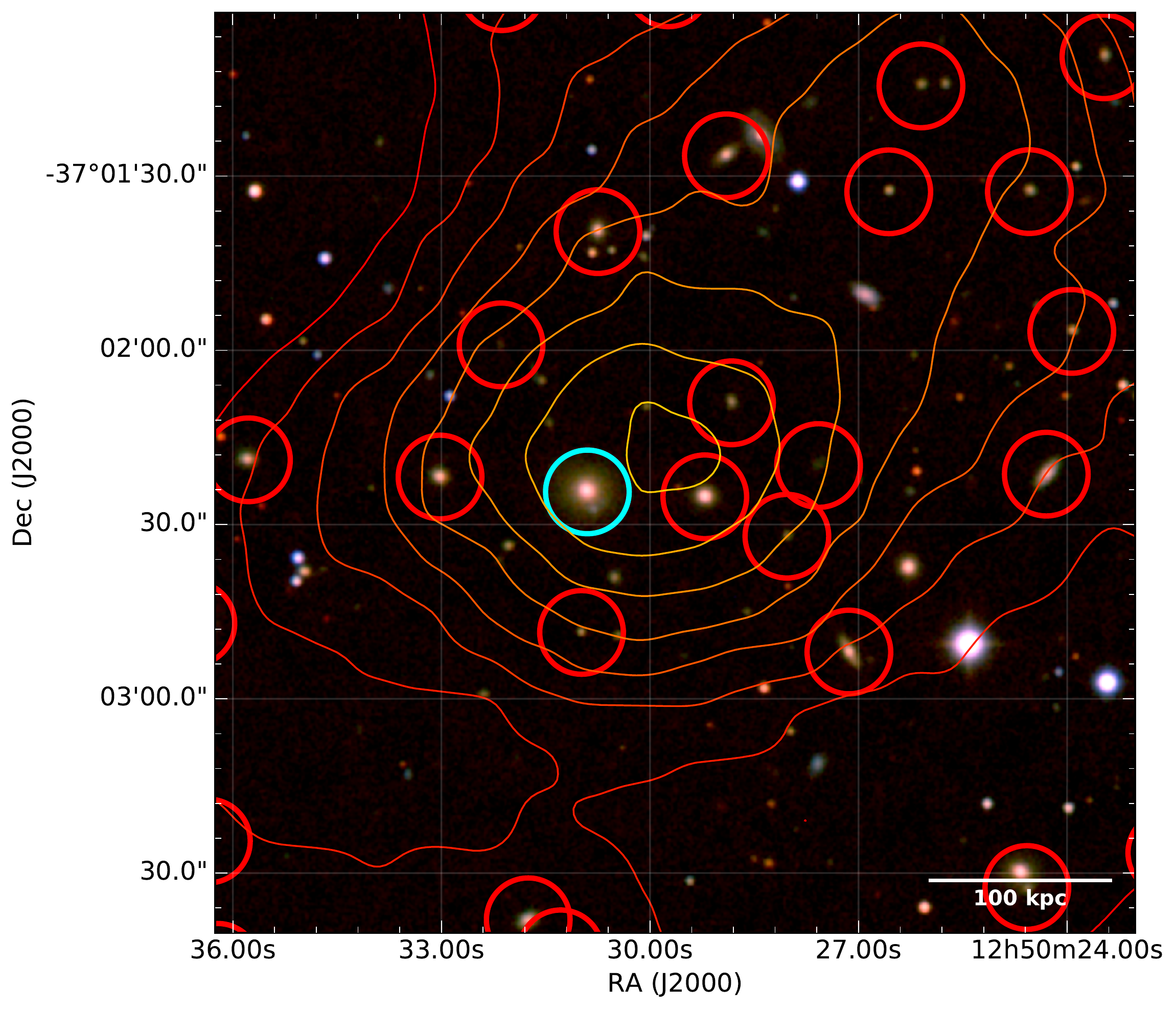}
\caption{Composite optical image ($g'r'z'$ filters) with photometric-confirmed cluster members marked with red circles, the brightest cluster galaxy (BCG) is marked with a cyan circle. Contours trace the galaxy density distribution weighted by the galaxy \texttt{r}-band magnitude. The position of the field of view with respect to the cluster is shown in Fig.\ref{fig:rosat}. On the right panel we show a zoom on the central part of the field.}\label{fig:opt}
\end{figure*}

\begin{figure}[!ht]
\centering
\includegraphics[width=.5\textwidth]{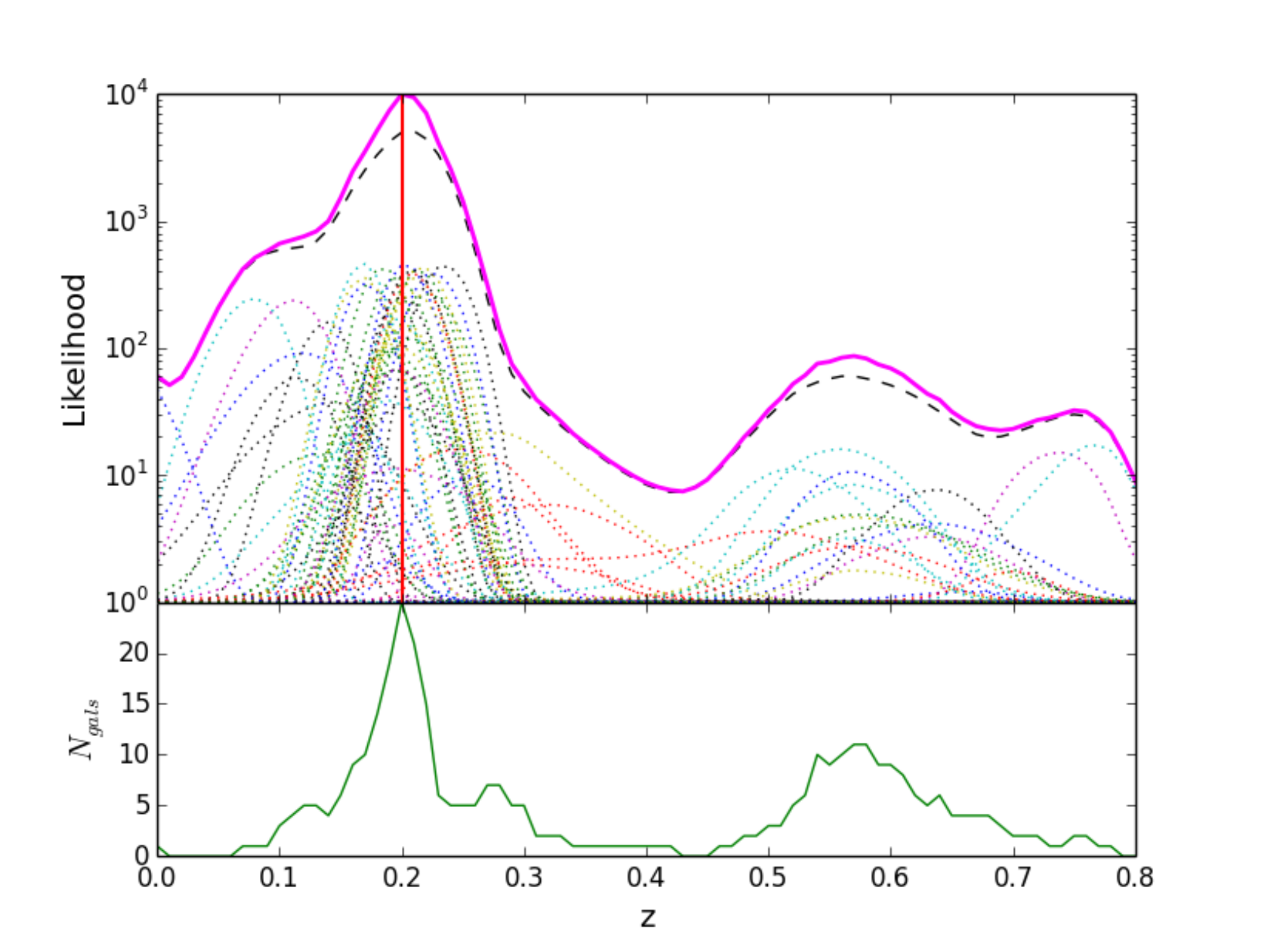}
\caption{Upper panel: The dotted lines show the likelihood distributions as a function of redshift for the individual galaxies. The black-dashed line is the sum of the individual galaxy distributions.  The solid magenta line is the product of the summed galaxy distributions multiplied by an indication of the cluster richness and gives the final likelihood distribution. These curves are arbitrarily normalised so that they can be plotted simultaneously.  Lower panel:  The number of galaxies that are most likely to be red sequence members at a given redshift.}\label{fig:photo}
\end{figure}

\subsubsection{Data reduction}
The total integration time of the observation was 460 s, split into four exposures of 115 s at different telescope dither patterns. During this observation the full-width-half-maximum was measured to be 0.85\arcsec in the $r$-band. Preliminary reduction of the data was performed using the methods of \cite{KupkuYoldas2008,Kruhler2008}. This pipeline is based on the standard tools of \textit{IRAF/PyRAF} and provides pixel and gain-corrected, astrometrised, stacked images for each channel. Some of the key details are summarised here.

A number of bias and dark frames were recorded directly following the observing night and flat-field observations were performed during twilight in the evening preceding the observations. Master bias, dark and flat-field frames were then produced by combining the individual exposures applied to the data. A master fringe pattern for the $i$ and $z$ bands was created by combining those generated for each of the four individual exposures. This master pattern was then subtracted from each frame individually before they were combined into the final coadd for each filter, and finally, the sky background was calculated from each of these coadds with the sources masked out and subtracted from the image.

An initial astrometric solution was found through the matching of stars in common with the USNO-A2.0 catalogue for the optical bands and the 2MASS catalogue \citep{Skrutskie2006} for the near-IR bands. This astrometric solution was refined by making use of the publicly available software \texttt{SCAMP} \citep{Bertin2006} and the coadded images in the respective bands mapped to a common pixel grid with the use of \texttt{SWARP} \citep{Bertin2002}, a publicly available software that performs the resampling and co-addition of FITS images.

A general model for the PSF across the field-of-view was constructed by making use of bright, unsaturated stars and by making use of the publicly available software \texttt{PSFEx} \citep{Bertin2011}. Source detection and photometric measurements were performed using \texttt{SExtractor}, operating in dual-mode with the $r$-band as the detection image. Star-galaxy separation was accomplished by selecting objects based on the \texttt{SExtractor} parameters CLASS\_STAR and SPREAD\_MODEL for the $r$-band. Photometric calibration was performed via observation of standard stars in the Sloan Digital Sky Survey (SDSS) and the zero-point corrections to the GROND photometry were determined by comparing PSF magnitudes in the two catalogues. These zero-point corrections were then applied to the observation of Abell 3527-bis. Corrections for galactic extinction were then applied to the GROND object magnitudes based on the dust maps of \cite{Schlegel1998} and making use of the Python package Astroquery.  \texttt{SExtractor} Kron magnitudes, MAG\_AUTO, were chosen for the calculation of galaxy colours. Sources were detected in the inner 4.5\arcmin$\times$4.5\arcmin of the pixel matched images.

\subsubsection{Photometric redshifts}
The approach to determining cluster photometric redshifts is based on the colour-magnitude redshift technique and follows a simplified version of the methodology of Ridl et al. (2016, in prep.), where GROND observations were utilised to determine photometric redshifts of X-ray selected galaxy clusters in the redshift range $0<z<0.8$. The exposure time chosen in Ridl et al. (2016, in prep.) for the redshift range $0<z<0.3$ matches that used for the data presented here and the same combination of filters (\texttt{g'r'i'z'}) is used in the photometry redshift calculations. All of these indicate that this analysis is sufficient to confirm the detection of a cluster and provide a good estimate of the photometric redshift. The key points are summarised below.

First, galaxies were selected from the \texttt{SExtractor} catalogues by selecting those with CLASS\_STAR $< 0.7$, $r$-band magnitude less than 24.0 and signal-to-noise in the Kron aperture greater than 5.0. For each galaxy, the probability of it being a cluster red sequence member at each redshift in the range $0<z<1.1$ in increments of $\delta z = 0.01$ was calculated according to its distance in colour-space of $g-r$, $r-i$, $i-z$, assuming a Gaussian distribution with a width of 0.05. The sum over all galaxies was then calculated giving a likelihood distribution for the entire cluster and multiplied by the number of galaxies with a high enough contribution to the cluster at each redshift. This is expressed in Equation (\ref{gaussian}) below:
\begin{equation}\label{gaussian}
p(z) = N_{gal}(z)\sum\limits_{gal}\left\{ \prod\limits_{c} \frac{1}{\sqrt{2\pi}\sigma}\exp\left[\frac{-(c_{gal} - c_{model})^2}{\sigma_c^2}\right]\right\}.
\end{equation}
The product runs over all colour combinations $[g-r, r-i, i-z]$. $\sigma_c = \sqrt{0.05^2+\sigma_{c,phot}^2}$ combines the width of the red sequence and the error on the photometry. $c_{gal}$ is the galaxy colour,  $c_{model}$ is the model colour and $N_{gal}(z)$ is the number of galaxies with a significant contribution to the cluster likelihood at redshift $z$. The expected colour, $c_{model}$, is derived by convolving the GROND transmission curves with the expected spectral energy distribution (SED) of a typical elliptical galaxy using the model provided by \cite{Polletta2007} and making use of \texttt{LePhare} \citep{Ilbert2006}. Based on the resulting likelihood distribution we find a clear peak at a photometric redshift of $z=0.20\pm0.02$. A total of 44 galaxies were found to be part of the cluster within a radius of $\sim 0.5$~Mpc.

\section{\target{}: a new galaxy cluster}
\label{sec:target}

\begin{figure}
\centering
\includegraphics[width=.49\textwidth]{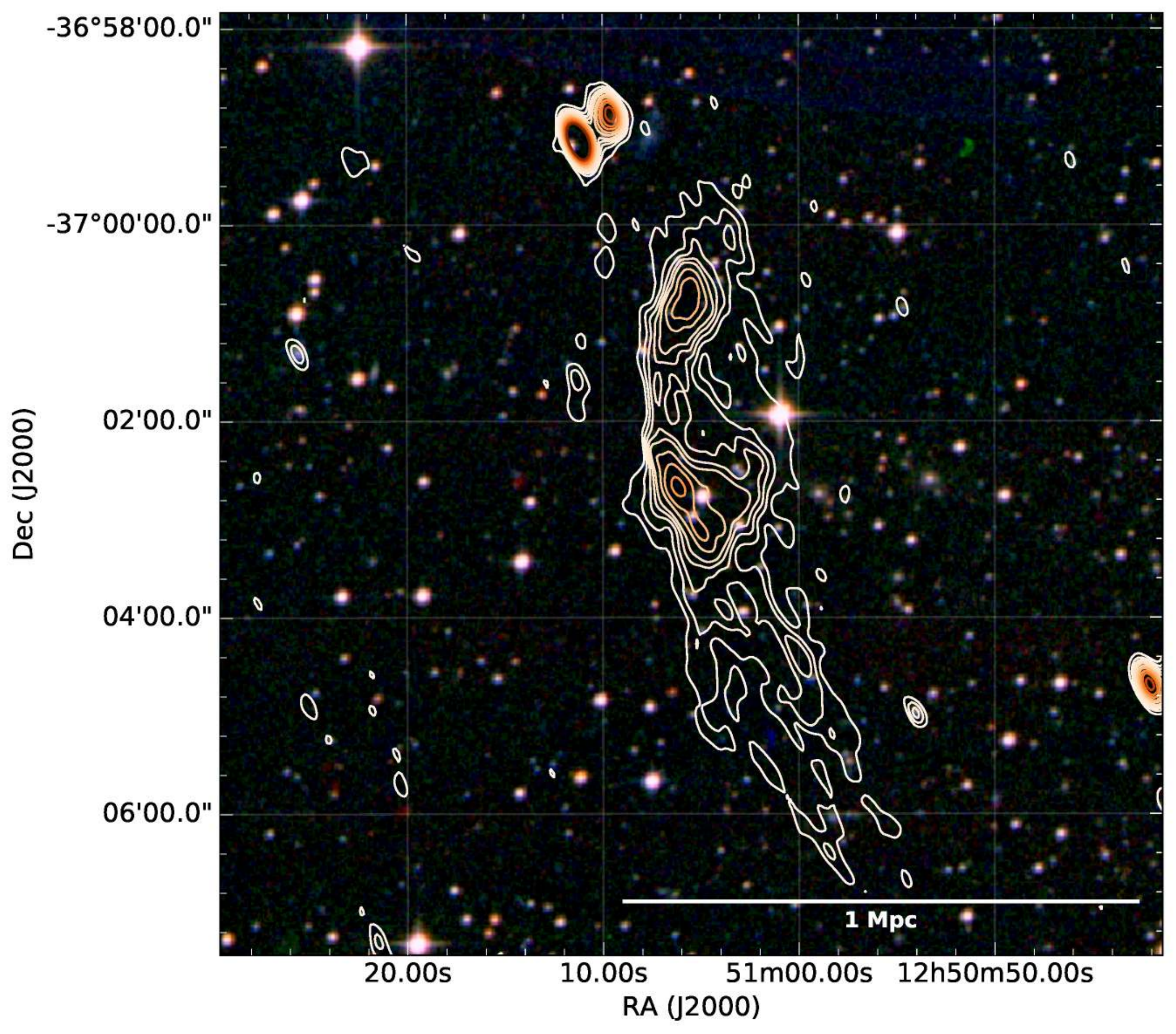}
\caption{Composite DSS image (\texttt{b}, \texttt{r}, and \texttt{i} filters) of the radio-relic region with radio contour superimposed. Contours are the same as in Fig.~\ref{fig:325}.}\label{fig:opt-relic}
\end{figure}

The outcome of radio observations, shown in Fig.~\ref{fig:radio}, confirms the presence of a spectacular giant radio relic. The radio relic in \target{} presents several irregularities compared to other examples \citep[e.g. the ``Sausage'' cluster;][]{vanWeeren2010a}. However, as expected for radio relics, the source is slightly curved with the concave side facing the cluster. The flux density on the side of the relic facing the cluster mildly decreases (likely due to synchrotron and inverse Compton looses) while the other side of the source, where particles are accelerated, shows the sharper edge. No sign of a counter-relic is detected at any frequency.

\begin{figure}
\centering
\includegraphics[width=.49\textwidth]{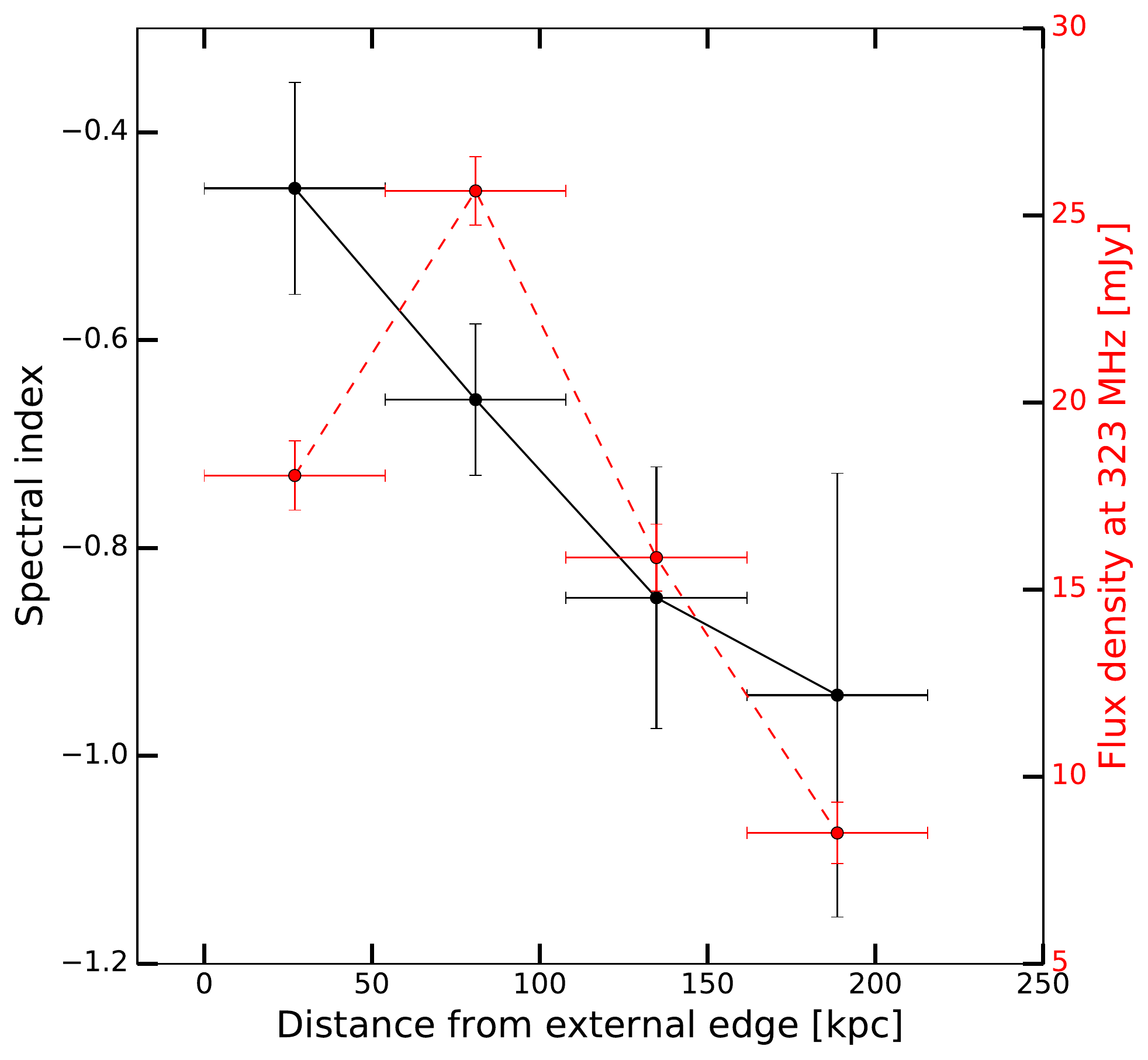}
\caption{Black solid-line: radio spectral index between 323 and 608 MHz. Red dashed-line: radio flux density at 323 MHz across the source extension. The data-points correspond to the regions displayed in Fig.~\ref{fig:spidx}. Error-bars on the x-axes correspond to the region size.}\label{fig:spidx-plot}
\end{figure}

As already seen in other radio relics \citep{Bonafede2009a}, the spectral radio-relic index map (see Fig.~\ref{fig:spidx}) is quite irregular. As radio relics are generated by outward travelling shock-waves, the electrons are accelerated on the outer edge of the relic and their spectra steepen by synchrotron and inverse Compton losses moving inward. As a consequence a strong sign that a radio source is a radio relic is a steepening spectrum along the minor axis of the source moving from outside-in with respect to the galaxy cluster. We tested this hypothesis by drawing four beam-sized, curved regions along the relic extensions and by calculating the average spectral index inside these regions. The results are plotted in Fig.~\ref{fig:spidx-plot}, where we show that the spectral index steepens along the relic.

The radio-relic classification of the radio source is based on several characteristics commonly present in this type of sources: (i) its morphology as discussed above, (ii) it is $\sim1.3$ Mpc away from the peak of the X-ray emission, (iii) its largest linear size is $\sim 1-1.5$ Mpc, (iv) it does not show any compact-bright radio emission or optical counterpart (see Fig.~\ref{fig:opt-relic}), and (v) its spectral index decreases towards the cluster centre (see Fig.~\ref{fig:spidx-plot}). The most important properties of the galaxy cluster and associated radio relics are listed in Table~\ref{tab:cluster}. If the source was a radio galaxy with an undetected far counterpart, we would expect it to be extremely large (likely $>$~few~Mpc), with no radio core, and strongly distorted in a very peculiar way. We believe that the radio-relic classification is much more realistic.

The ROSAT X-ray emission from \target{} was (blindly) associated with a galaxy at $z\simeq0.19$ \citep{Mahony2010}. With a pointed optical observation we confirmed the presence of a newly discovered galaxy cluster co-located with the X-ray excess. The magnitude-weighted galaxy density distribution of the cluster appears elongated in the same direction of the X-ray emission (see Fig.~\ref{fig:opt}). We note, however, that a larger number of galaxies on a larger field of view are required to confirm this claim. The galaxy found by \cite{Mahony2010} is actually the BCG of this galaxy cluster. The X-ray excess is 20\arcmin{} away from a known galaxy cluster (Abell 3527, $z=0.1983$). This means that \target{} is $\sim4$~Mpc away from Abell 3527 at about the same redshift, creating a rare system of two clusters at close distance.

\begin{table*}
\centering
\begin{threeparttable}
\begin{tabular}{lc}
\hline
Right ascension\tnote{a} & \hms{12}{50}{30.900}\\
Declination\tnote{a} & \dms{-37}{02}{24.39}\\
Redshift & $0.20\pm0.02$ \\
X-ray luminosity (0.5--2 kev) & $(1.5 \pm 0.2) \times 10^{44}$ erg/s\\
X-ray luminosity (0.1--2.4 kev) & $(1.9^{+0.3}_{-0.4}) \times 10^{44}$ erg/s\\
$M_{500}$ & $(3.3\pm0.3) \times 10^{14}$~\Msun\\
Relic flux density (148 MHz) & $277 \pm 12$~mJy \\
Relic flux density (323 MHz) & $98 \pm 1$~mJy\\
Relic flux density (608 MHz) & $70 \pm 1$~mJy\\
Relic flux density\tnote{b}\ \ (1400 MHz) & $35 \pm 2$~mJy\\
Relic luminosity\tnote{c}\ \ (323 MHz) & $(1.07 \pm 0.01) \times 10^{32}$ erg/s/Hz\\
Relic luminosity\tnote{bc}\ \ \ (1400 MHz) & $(3.83 \pm 0.03 ) \times 10^{31}$ erg/s/Hz\\
Relic integrated spectral index & $-0.70 \pm 0.04$\\
Relic size & 1.0--1.5 Mpc\\
Relic distance from X-ray peak & 1.3 Mpc\\
\hline
\end{tabular}
\begin{tablenotes}
    \item[a] Coordinates of the BCG.
    \item[b] Rescaled with spectral index -0.7 from 323 MHz.
    \item[c] k-corrected with spectral index -0.7.
\end{tablenotes}
\end{threeparttable}
\caption{Galaxy cluster properties}\label{tab:cluster}
\end{table*}

\section{Discussion}
\label{sec:discussion}
A prediction from simple DSA theory is a relation between the Mach number of the shock and the injection spectral index of the accelerated electrons:
\begin{equation}\label{eq:mach}
 \mach = \sqrt{\frac{2\alpha_{\rm inj}-3}{2\alpha_{\rm inj}+1}}.
\end{equation}
Therefore, a spectral index $>-0.5$ implies an unphysical Mach number. In the case of this specific relic the injection spectral index appears extremely flat. If we assume the outer region as representative for the injection index we find: $\alpha_{\rm inj} = -0.45 \pm 0.10$. This value seems to point towards a stronger Mach number shock than what is usually found for radio relics. However, we note that Eq.~\ref{eq:mach} has been shown not to hold in at least the well-studied case of the ``toothbrush'' cluster \citep{vanWeeren2016}. Also the integrated spectral index (extracted from $uv$-harmonised images at 323 and 608 MHz) appears quite flat ($\alpha = -0.70 \pm 0.04$) with respect to the average spectral index of radio relics \citep[$\alpha = -1$;][]{Feretti2012}. As already noted for another radio relic with a flat integrated spectral index \citep[Abell 2256;][]{vanWeeren2012,Trasatti2015}, if $\alpha > -1$ the stationary shock conditions do not hold. As a consequence, a possible way to reconcile the flat spectrum with shock acceleration is that the relic must have been produced very recently ($\sim0.1$ Gyr ago). This means that the time for energy losses is longer than the age of the shock at some observed energies and that the integrated spectrum of the relic will still evolve, getting steeper. We note that using the total flux density at 148, 323, 608, and 1400 MHz ($28\pm3$~mJy; from NVSS), we obtain a steeper integrated spectral index of $\alpha = -0.98 \pm 0.04$. However, given the different $uv$-coverages, these maps can be sensitive to different spatial scales.

The X-ray luminosity extracted from ROSAT data is $L_{\rm [0.5-2 {\rm kev}]} = (1.2 \pm 0.2) \times 10^{44}$~erg/s. From this luminosity, we derived the associated cluster mass: $M_{500} \simeq 3.3\pm0.3 \times 10^{14}$~\Msun{} \citep{Pratt2009}. The cluster is amongst the least massive clusters hosting a radio relic discovered so far \citep[see Fig.7b in][]{deGasperin2014c}. Only two other less-massive systems are known to host radio relics: Abell~3365 \citep[$z=0.0926$, $M_{500} \simeq 2.00\pm0.08 \times 10^{14}$~\Msun{}, from X-ray;][]{vanWeeren2011} and Abell~3376 \citep[$z=0.046$, $M_{500} \simeq 2.3\pm0.2 \times 10^{14}$~\Msun{}, from SZ;][]{Kale2012a}. Both galaxy clusters host a double radio-relic system. In \cite{deGasperin2014c}, we discussed the presence of a scaling relation between radio-relic power and hosting-cluster mass. Given that relation, \target{} is almost an order of magnitude over-luminous for the hosting cluster mass. Nonetheless, \target{} provides a rare example of low-mass cluster hosting a radio relic at $z>0.1$.

\cite{Nuza2012} investigated the predicted abundance of radio relics in current and future radio surveys. They estimated the probability of generating a radio relic by running cosmological simulations. Then they normalised the predicted radio-relic number counts using the bright end of the currently observed number-count distribution. According to their results, a radio relic such as the one described in this paper, with a flux density $S_{\rm 1.4 GHz} \sim35$~mJy in a cluster with $L_{X} < 3 \times 10^{44}$ erg/s, is extremely unlikely. This can rely on the fact that \cite{Nuza2012} assume a radio-relic power--cluster mass correlation to predict the number of observable relics and this relic is largely overluminous with respect to this correlation; thus, simply it cannot be predicted on the basis of their assumption. Another possibility could be that we are underestimating the cluster X-ray luminosity of \target{}, or that we are dealing with an intrinsically peculiar high-power relic. They also provide an estimation of the fraction of clusters with a detectable radio relic for a given X-ray luminosity \citep[Fig.\,8,][]{Nuza2012}. At the luminosity of \target{} (properly rescaled in their energy band) the probability of finding a radio relic is $<{\rm few}\%$ with current instrumentation. \cite{Nuza2012} also estimate that with the LOFAR sky survey this fraction will rise to $~\sim 50\%$. However, they assumed that a detection of a radio relic is possible only when the cluster position is known a priori. As an alternative, the combined information on the source morphology, spectral index and polarisation, might be able to reverse the paradigm so that radio relics will be used to trace galaxy clusters, as in the case described in this paper.

\section{Conclusions}
\label{sec:conclusions}
We reported the discovery of a radio relic located in the outskirts of a previously unknown galaxy cluster: \target{} ($z=0.20\pm0.02$). The relic nature of the source has been confirmed from morphological and spectral properties. The most interesting characteristic of this source is its high luminosity and unusually flatter spectral index which, assuming simple DSA prescription, corresponds to a rather high Mach number in the associated shock wave.

The associated galaxy cluster is also peculiar for its low mass ($M_{500}=3.3\pm0.3 \times 10^{14}$~\Msun). This is the third least massive system known to date to host a radio relic. However, better X-ray data are required to firmly assess the global cluster characteristics. Furthermore, together with the extremely bright ($S_{\rm 1.4\ GHz} = 0.32 \pm 0.02$~Jy) radio relic in the ``toothbrush'' cluster \citep{VanWeeren2012e}, \target{} is the only other galaxy cluster to have been detected thanks to the presence of a radio relic. This detection proves that radio surveys can be used to locate a subset of relatively low mass merging galaxy clusters ($M_{500} \simeq {\rm few} \times 10^{14}$~\Msun) and that the occurrence of radio relics in low mass clusters may be more common than previously thought.

\begin{acknowledgements}

We would like to thank the staff of the GMRT that made these observations possible. GMRT is run by the National Centre for Radio Astrophysics of the Tata Institute of Fundamental Research.







Based on data obtained with the MPG 2.2 m telescope at the ESO La Silla Observatory. Part of the funding for GROND (both hardware as well as personnel) was generously granted from the Leibniz-Prize to Prof. G. Hasinger (DFG grant HA 1850/28-1).

\end{acknowledgements}


\bibliographystyle{aa}
\bibliography{../bbtex/papers-archetto}


\onecolumn
\begin{appendix}

...

\end{appendix}


\end{document}